\documentstyle[12pt,aasms4]{article}

\begin{document}

\title{VERY HIGH ENERGY GAMMA RAYS FROM THE VELA PULSAR DIRECTION}

\author{T.~Yoshikoshi\altaffilmark{1,2}, T.~Kifune\altaffilmark{2},
  S.A.~Dazeley\altaffilmark{3}, P.G.~Edwards\altaffilmark{3,4},
  T.~Hara\altaffilmark{5}, Y.~Hayami\altaffilmark{1},
  F.~Kakimoto\altaffilmark{1}, T.~Konishi\altaffilmark{6},
  A.~Masaike\altaffilmark{7}, Y.~Matsubara\altaffilmark{8},
  T.~Matsuoka\altaffilmark{8}, Y.~Mizumoto\altaffilmark{9},
  M.~Mori\altaffilmark{10}, H.~Muraishi\altaffilmark{11},
  Y.~Muraki\altaffilmark{8}, T.~Naito\altaffilmark{12},
  K.~Nishijima\altaffilmark{13}, S.~Oda\altaffilmark{6},
  S.~Ogio\altaffilmark{1}, T.~Ohsaki\altaffilmark{1},
  J.R.~Patterson\altaffilmark{3}, M.D.~Roberts\altaffilmark{2},
  G.P.~Rowell\altaffilmark{3}, T.~Sako\altaffilmark{8},
  K.~Sakurazawa\altaffilmark{1}, R.~Susukita\altaffilmark{7,14},
  A.~Suzuki\altaffilmark{6}, T.~Tamura\altaffilmark{15},
  T.~Tanimori\altaffilmark{1}, G.J.~Thornton\altaffilmark{2,3},
  S.~Yanagita\altaffilmark{11} and T.~Yoshida\altaffilmark{11}}

\altaffiltext{1}{Department of Physics, Tokyo Institute of Technology,
  Meguro-ku, Tokyo 152, Japan}
\altaffiltext{2}{Institute for Cosmic Ray Research, University of Tokyo,
  Tanashi, Tokyo 188, Japan}
\altaffiltext{3}{Department of Physics and Mathematical Physics, University of
  Adelaide, Adelaide, South Australia 5005, Australia}
\altaffiltext{4}{Institute of Space and Astronautical Science, Sagamihara,
  Kanagawa 229, Japan}
\altaffiltext{5}{Faculty of Management Information, Yamanashi Gakuin
  University, Kofu, Yamanashi 400, Japan}
\altaffiltext{6}{Department of Physics, Kobe University, Kobe, Hyogo 637,
  Japan}
\altaffiltext{7}{Department of Physics, Kyoto University, Kyoto, Kyoto 606,
  Japan}
\altaffiltext{8}{Solar-Terrestrial Environment Laboratory, Nagoya University,
  Nagoya, Aichi 464, Japan}
\altaffiltext{9}{National Astronomical Observatory of Japan, Mitaka,
  Tokyo 181, Japan}
\altaffiltext{10}{Department of Physics, Miyagi University of Education,
  Sendai, Miyagi 980, Japan}
\altaffiltext{11}{Faculty of Science, Ibaraki University, Mito, Ibaraki 310,
  Japan}
\altaffiltext{12}{Department of Earth and Planetary Physics, University of
  Tokyo, Bunkyo-ku, Tokyo 113, Japan}
\altaffiltext{13}{Department of Physics, Tokai University, Hiratsuka,
  Kanagawa 259, Japan}
\altaffiltext{14}{Institute of Physical and Chemical Research, Wako,
  Saitama 351-01, Japan}
\altaffiltext{15}{Faculty of Engineering, Kanagawa University, Yokohama,
  Kanagawa 221, Japan}

\authoremail{tyoshiko@icrhp9.icrr.u-tokyo.ac.jp}

\begin{abstract}
We have observed the Vela pulsar region at TeV energies using the 3.8~m
imaging \v{C}erenkov telescope near Woomera, South Australia between
January 1993 and March 1995. Evidence of an unpulsed gamma-ray signal
has been detected at the 5.8~$\sigma$ level. The detected gamma-ray flux is
$(2.9 \pm 0.5 \pm 0.4) \times 10^{-12}\; \rm photons\; cm^{-2}\; sec^{-1}$
above $2.5 \pm 1.0$~TeV and the signal is consistent with steady emission
over the two years. The gamma-ray emission region is
offset from the Vela pulsar position to the southeast by about $0\fdg13$.
No pulsed emission modulated with the pulsar period has been detected and
the 95~\% confidence flux upper limit to the pulsed emission from the pulsar
is $(3.7 \pm 0.7) \times 10^{-13}\; \rm photons\; cm^{-2}\; sec^{-1}$
above $2.5 \pm 1.0$~TeV.
\end{abstract}


\section{Introduction}

It is highly likely that high energy particles are accelerated by
non-thermal processes in the vicinity of pulsars,
and extensive searches have been made for very high energy (VHE;
typically at the TeV energy) gamma rays from this class of objects 
(Kifune 1996\markcite{kifu1996}; references therein).
Positive evidence has been obtained for the Crab
(e.g. Weekes {\it et al.} 1989\markcite{week1989};
Vacanti {\it et al.} 1991\markcite{vaca1991};
Tanimori {\it et al.} 1994\markcite{tani1994})
and PSR~B$1706-44$ (Kifune {\it et al.} 1995\markcite{kifu1995})
using the technique of imaging \v{C}erenkov light from extensive air
showers which VHE gamma rays initiate in the upper atmosphere.
The Vela pulsar, PSR~B$0833-45$, is one of the 100~MeV gamma-ray pulsars
detected by EGRET of the {\it Compton Gamma Ray Observatory},
together with the Crab and PSR~B$1706-44$
(e.g. Ramanamurthy {\it et al.} 1995\markcite{rama1995}).

There have been some claims for VHE gamma-ray emission from the Vela pulsar
so far (Grindlay {\it et al.} 1975\markcite{grin1975}; Bhat {\it et al.}
1987\markcite{bhat1987}). However, no positive evidence has been found
in later observations (Bowden {\it et al.} 1993\markcite{bowd1993};
Nel {\it et al.} 1993\markcite{nel1993};
Edwards {\it et al.} 1994\markcite{edwa1994}).
In these observations, period analyses were used to reduce cosmic-ray
background counts and to search for pulsed gamma-ray emission. 
On the contrary, VHE emission from the Crab and PSR~B$1706-44$ is
apparently unpulsed. Therefore, it is important to look for
the unpulsed VHE gamma-ray signal also from the Vela pulsar.
For the Crab pulsar, intense X-ray
emission is observed as an X-ray synchrotron nebula of which the inverse
Compton counterpart is the source of VHE gamma rays. The energy injection,
which is close to the spindown luminosity of the Crab pulsar (PSR~B$0531+21$),
is through a flow of relativistic electrons and positrons, i.e.
a ``pulsar wind''(Kennel \& Coroniti 1984\markcite{kenn1984}).
The shock generated when the pulsar wind collides with circumstellar
matter is likely to be the site of particle acceleration.
It is likely that a similar process takes place at the Vela pulsar.

The Vela pulsar is relatively close to us, at a distance of about 500~pc,
and we can expect to resolve the structure of the pulsar wind and nebula
better than in the case of the Crab (at a distance of $\sim 2$~kpc).
In the X-ray energy band, a compact nebula was detected by the {\it Einstein}
satellite with fainter extended emission
(Harnden {\it et al.} 1985\markcite{harn1985}),
which was later identified as an X-ray jet by the {\it ROSAT} satellite
extending from the pulsar to the south-southwest direction and which
Markwardt \& \"{O}gelman (1995)\markcite{mark1995} argued may be
evidence of a pulsar wind from the Vela pulsar. The ``head'' of the Vela jet
has been observed also by the {\it ASCA} satellite and the authors explored
the possibility that the jet emission is non-thermal (synchrotron) radiation
rather than thermal (Markwardt \& \"{O}gelman 1997\markcite{mark1997}).
If the X-ray emission is of non-thermal origin, the detection of
Compton-boosted VHE gamma rays would provide direct and clear evidence
of non-thermal electrons.

\section{Instruments and Observations}

The 3.8~m telescope of the CANGAROO collaboration 
(Hara {\it et al.} 1993\markcite{hara1993}) detects \v{C}erenkov photons
from extensive air showers generated by primary gamma rays or cosmic rays,
and has an imaging camera of 256 Hamamatsu R2248 square photomultiplier tubes.
One photomultiplier tube views an angular extent of $0\fdg12 \times 0\fdg12$,
and the total field of view of the camera is about $3\arcdeg$ across.

The tracking accuracy of the telescope has been checked by monitoring
the tracks of bright stars in the field of view. 
Monte Carlo simulations show the accuracy of calibrating the target position
using star tracks to be better than $0\fdg02$
(Yoshikoshi 1996\markcite{yosh1996}).
At least four bright stars of visual magnitude $5 \sim 6$ are in the 
$3\arcdeg$ field of view around the Vela pulsar and the 
telescope pointing has been checked night by night.

The Vela pulsar has been observed by the 3.8~m telescope from January 1993
to March 1995. The on-source data amount to about 174~hours in total
with a cosmic-ray event rate of about 1~Hz. Almost the same amount of
off-source data has been taken, in paired on-source and off-source runs
each night. About 119~hours usable on-source data remain after rejecting
the data affected by clouds. Only the data taken without cloud are used
in the following analyses.

\section{Analyses and Results}

The \v{C}erenkov imaging analysis is based on
parameterization of the shape, location and orientation of approximately
elliptical \v{C}erenkov images detected by an imaging camera.
The commonly used parameters are ``width'', ``length'', ``conc'' (shape),
``distance'' (location) and ``alpha'' (orientation)
(Hillas 1985\markcite{hill1985}; Weekes {\it et al.} 1989\markcite{week1989};
Reynolds {\it et al.} 1993\markcite{reyn1993}) and
the gamma-ray selection criteria in the present analysis are
$0\fdg01 < {\rm width} < 0\fdg09$,
$0\fdg1 < {\rm length} < 0\fdg4$, $0.3 < {\rm conc}$,
$0\fdg7 < {\rm distance} < 1\fdg2$ and ${\rm alpha} < 10\arcdeg$.
Monte Carlo simulations show that more than 99~\% of background events
are rejected after the above selections are made, while $\sim 50$~\% of
gamma-ray images from a point source remain.

First, we searched for a gamma-ray signal from the pulsar position
and found a significant excess of on-source events above the background
(off-source) level. However, the peak profile of the excess distribution
against the orientation angle ``alpha'' was broader than expected from
a point source. Next, the assumed position of gamma-ray emission
was shifted from the pulsar position, to scan a $2\arcdeg \times 2\arcdeg$
area around the pulsar within the field of view.
The position of the maximum excess counts
was searched for and found to be offset from the pulsar by about $0\fdg13$.
Figure~1 shows the alpha distributions for the position of
the maximum excess counts after the other selection criteria have been
applied.
\placefigure{fig1}
The statistical significance of the excess after including the selection
of alpha is at the 5.2~$\sigma$ level, where the statistical significance
is calculated by
$(N_{\rm on} - N_{\rm off}) / \sqrt{N_{\rm on} + N_{\rm off}}$,
and $N_{\rm on}$ and $N_{\rm off}$ are the numbers of gamma-ray-like events
in the on-source and off-source data, respectively.

In the usual image analysis, the image parameters are calculated from
the mean and second moments of light yield, i.e. assuming gamma-ray images
to have a simple elliptical shape. Gamma-ray images, however,
are asymmetrical along their major axes, having an elongated, 
comet-like shape with the ``tail'' pointing away from the source position.
This feature can be characterized and quantified by another image parameter
called the ``asymmetry'', which is the cube root of the third moment
of the image along the major axis normalized by ``length''
(Punch 1993\markcite{punc1993}). Figure~2~(a) shows the asymmetry
distributions for gamma-ray images for a point source and background
(proton) images, inferred from Monte Carlo simulations.
\placefigure{fig2}
About 80~\% of gamma-ray images have positive values of ``asymmetry''
in the simulation, while the distribution of background images is
almost symmetrical because of their isotropic arrival
directions. The asymmetry analysis has been applied assuming the source
lies at the position of the maximum excess counts,
to examine the gamma-ray-like feature of the excess events.
The on-source distribution appears nearly symmetrical as shown in Figure~2~(b).
However, after subtracting the off-source distribution, the distribution
of the excess events in Figure~2~(c) is asymmetrical with the peak
on the positive side as expected from the simulation. This result gives
clear corroborative evidence that the detected excess events of Figure~1
are truly due to gamma rays. The significance of 5.2~$\sigma$
for the excess counts in the alpha distribution of Figure~1 increases
to 5.8~$\sigma$ by adding ${\rm asymmetry} > 0$ to the gamma-ray selection
criteria.

As an alternative method to calculating the alpha parameter, it is possible
to infer and assign the true direction of a gamma ray from the location and
orientation of its observed \v{C}erenkov image. We have calculated
a probability density for the true direction for each observed
gamma-ray image using Monte Carlo simulations
(Yoshikoshi 1996\markcite{yosh1996}). Thus, a density map
of gamma-ray directions can be obtained by adding up all of the probability
densities of the gamma-ray-like images. The contribution of
the gamma-ray source in the field of view can then be found as
an event excess in the on-source map over the background (off-source) map.
Figure~3 shows the density map of the excess events
for the Vela pulsar data plotted as a function
of right ascension and declination.
\placefigure{fig3}
The significant excess due to a gamma-ray source near the pulsar exists
in this map. The position of maximum emission is offset from the pulsar,
which is indicated by the ``star'' mark in the map, to a position southeast
by about $0\fdg13$. The alpha distribution of Figure~1 is plotted for
the position of maximum emission.

The gamma-ray integral flux calculated for the position of
the maximum excess counts is
$(2.9 \pm 0.5 \pm 0.4) \times 10^{-12}\; \rm photons\; cm^{-2}\; sec^{-1}$
above 2.5~TeV, where the first and second errors are statistical and
systematic respectively. The flux from the pulsar position is
$1.4 \times 10^{-12}\; \rm photons\; cm^{-2}\; sec^{-1}$ above 2.5~TeV.
However, the excess counts which are calculated for the positions of
the maximum excess counts and the pulsar are not independent of each other
and all of the flux measured from the pulsar position is possibly
contamination by the events from the position of the maximum excess counts.
The threshold energy is defined as the energy of the maximum differential
flux of detected gamma-ray-like events and has been estimated from Monte Carlo
simulations to be 2.5~TeV, assuming a power law spectrum with a photon index
of $-2.5$. The systematic error for this threshold
energy is about $\pm 1.0$~TeV, which is due to uncertainties in
the photon index, the trigger conditions and the reflectivity of
the mirror. The data have been divided to calculate
year-by-year fluxes and the fluxes are consistent with no variation
having been detected on the time scale of two years.

The periodicity of the events detected from the Vela pulsar direction and
the direction of the maximum excess counts has been investigated
using our 1994 and 1995 data for which a global positioning system (GPS)
was available.
A derivative of the TEMPO programs (Taylor \& Weisberg 1989\markcite{tayl1989})
has been used to transform the event arrival times to
solar system barycentric (SSB) arrival times. The solar system ephemeris
used in the analysis comes from the Jet Propulsion Laboratory (JPL) and
is based on epoch 2000 (DE 200) (Standish 1982\markcite{stan1982}). We have
applied the $Z_2^2$ test (Buccheri {\it et al.} 1983\markcite{bucc1983})
to the data set after the gamma-ray selection with the period of
the Vela pulsar using ephemerides from the Princeton data base
(Arzoumanian {\it et al.} 1992\markcite{arzo1992}), and the test has
been done coherently for each year. No significant value of
the $Z_2^2$ statistic has been found for either the excess direction nor
the pulsar direction. The 95~\% confidence upper limit to the pulsed flux
has been calculated for the pulsar position as
$(3.7 \pm 0.7) \times 10^{-13}\; \rm photons\; cm^{-2}\; sec^{-1}$
above 2.5~TeV from the values of $Z_2^2$
using the method of Protheroe (1987)\markcite{prot1987}.

\section{Discussion}

The luminosity of VHE gamma rays from the position
of the maximum excess counts near the Vela pulsar is calculated to be
$\sim 6.0 \times 10^{32}\; \rm erg\; sec^{-1}$, assuming that the distance
to the gamma-ray emission region is 500~pc, the photon index of
the power law spectrum is $-2.5$, the maximum gamma-ray energy is
10~TeV and the emission is isotropic. The VHE luminosity corresponds
to $\sim 8.6 \times 10^{-5}$ of the pulsar's spindown energy of
$6.9 \times 10^{36}\; \rm erg\; sec^{-1}$, and is smaller by an order of
magnitude than the luminosity of pulsed emission in the range from 100~MeV
to 2~GeV (Kanbach {\it et al.} 1994\markcite{kanb1994}).
The luminosity of VHE gamma rays from the pulsar position is estimated to be
$\sim 2.9 \times 10^{32}\; \rm erg\; sec^{-1}$, which should rather be
interpreted as an upper limit because a majority of the excess events at
the pulsar position are common with the events of the maximum excess position.

The emission region of VHE gamma rays appears to differ by about $0\fdg13$
to the southeast from the Vela pulsar position. We have investigated
whether the effect of statistical fluctuations could cause a change
in the observed source position, using Monte Carlo simulations.
The position error derived for a point source of this strength is estimated
to be about $0\fdg04$, and so the observed offset of the detected peak
position from the pulsar corresponds to a significance of more than
3~$\sigma$. In this estimation, we ignored the effect due to
the source extent and, in fact, the peak profile which appears in Figure~3
seems to be rather more extended than the point spread function
(HWHM $\sim 0\fdg18$). We note also that separate maps of each year's data
show consistent offsets. The 3.8~m mirror has recently been recoated and
the Vela pulsar region has been observed with a better reflectivity
($\sim 90$~\%) again in 1997 further to confirm the offset. We have found
a gamma-ray-like event excess from the 1997 data with a significance of
more than 4~$\sigma$ and the position of maximum excess counts, indicated
by the ``cross'' in Figure~3, agrees with that from the 1993 to 1995 data
within the statistical error.

The southeast offset of $0\fdg13$ corresponds to about 1~pc for
the distance of 500~pc, much larger than the size of the light cylinder,
and thus, we can exclude the possibility that the detected signal is from
the pulsar magnetosphere or from the X-ray jet.
This conclusion is consistent with there being
no pulsation detected at the pulsar period.
Unpulsed VHE emission at a distance of $\sim 1$~pc from the pulsar
may suggest that the mechanism for generating relativistic particles and
inverse Compton gamma rays in the Vela pulsar and nebula is similar
to the case of the Crab pulsar and nebula.
A shock due to the confinement of the pulsar wind by the nebula may exist
or have formed at the position southeast of the Vela pulsar by about $0\fdg13$.
This position of the VHE emission agrees with the birthplace of the Vela pulsar
(\"{O}gelman {\it et al.} 1989\markcite{ogel1989};
Bailes {\it et al.} 1989\markcite{bail1989}). Accelerated electrons could
survive longer than the pulsar age if the magnetic field is not strong,
and the VHE emission may be due to such long-lived electrons as pointed out
by Harding \& de~Jager (1997)\markcite{hard1997}.
Recent X-ray observations show that a number of young pulsars are accompanied
by possible synchrotron nebulae (Kawai \& Tamura 1996\markcite{kawa1996}),
which are offset from the pulsars.
The existence of relativistic electrons in a region displaced from the pulsar
may be rather a common feature observed in pulsar/nebula systems.
VHE gamma rays could well be produced by electrons in such displaced regions
as we observe in the case of the Vela pulsar and nebula.

In the {\it ROSAT} X-ray image
(Markwardt \& \"{O}gelman 1995\markcite{mark1995}),
a ``bright spot'' can be found where we have detected
the VHE gamma-ray signal.
Assuming that the X-rays originate in synchrotron emission
from the same electrons that, through inverse Compton scattering,
produce the VHE gamma rays,
we can estimate the magnetic field from the relation
$L_{\rm syn} / L_{\rm iC} = U_B / U_{\rm ph}$
of inverse Compton luminosity $L_{\rm iC}$ and synchrotron luminosity
$L_{\rm syn}$, where $U_B = B^2 / 8 \pi$ and $U_{\rm ph}$ are the energy
densities of the magnetic field and the photon field, respectively.
The X-ray luminosity from the bright spot at the pulsar birthplace
is estimated from the {\it ROSAT} image to be smaller than that from
the $1\arcmin$ X-ray nebula around the pulsar,
$1.3 \times 10^{33}\; \rm erg\; sec^{-1}$ per decade of energy
(\"{O}gelman {\it et al.} 1993\markcite{ogel1993}),
while the VHE gamma-ray luminosity is
$\sim 1.0 \times 10^{33}\; \rm erg\; sec^{-1}$ per decade.
An upper limit for the magnetic field at the bright spot
is then estimated to be $B \lesssim 4 \times 10^{-6}
(U_{\rm ph} / 0.24\; {\rm eV\; cm^{-3}})^{1 / 2}$~G,
where $0.24\; \rm eV\; cm^{-3}$ is the energy density of the microwave
background photons.
As for the compact X-ray nebula,
a lower limit for the magnetic field can be estimated from the upper limit
for the VHE gamma-ray luminosity from the pulsar position, i.e.
$B \gtrsim 5 \times 10^{-6}
(U_{\rm ph} / 0.24\; {\rm eV\; cm^{-3}})^{1 / 2}$~G.
This limit is compatible with an estimate obtained by de~Jager {\it et al.}
(1996)\markcite{deja1996} from a confinement condition for the progenitor
electrons.
A plausible scenario is thus that the VHE gamma-ray emission occurs 
from the X-ray bright spot
offset $0\fdg13$ from the pulsar to the southeast where there is
sufficient density of relativistic electrons, but a smaller magnetic field.

\acknowledgments

This work is supported by International Scientific Research Program of
a Grant-in-Aid in Scientific Research of the Ministry of Education, Science,
Sports and Culture, Japan, and by the Australian Research Council.
T.~Kifune and T.~Tanimori acknowledge the support of the Sumitomo Foundation.
The receipt of JSPS Research Fellowships (P.G.E., T.N., M.D.R., K.S., G.J.T.
and T.~Yoshikoshi) is also acknowledged.

\clearpage

\figcaption[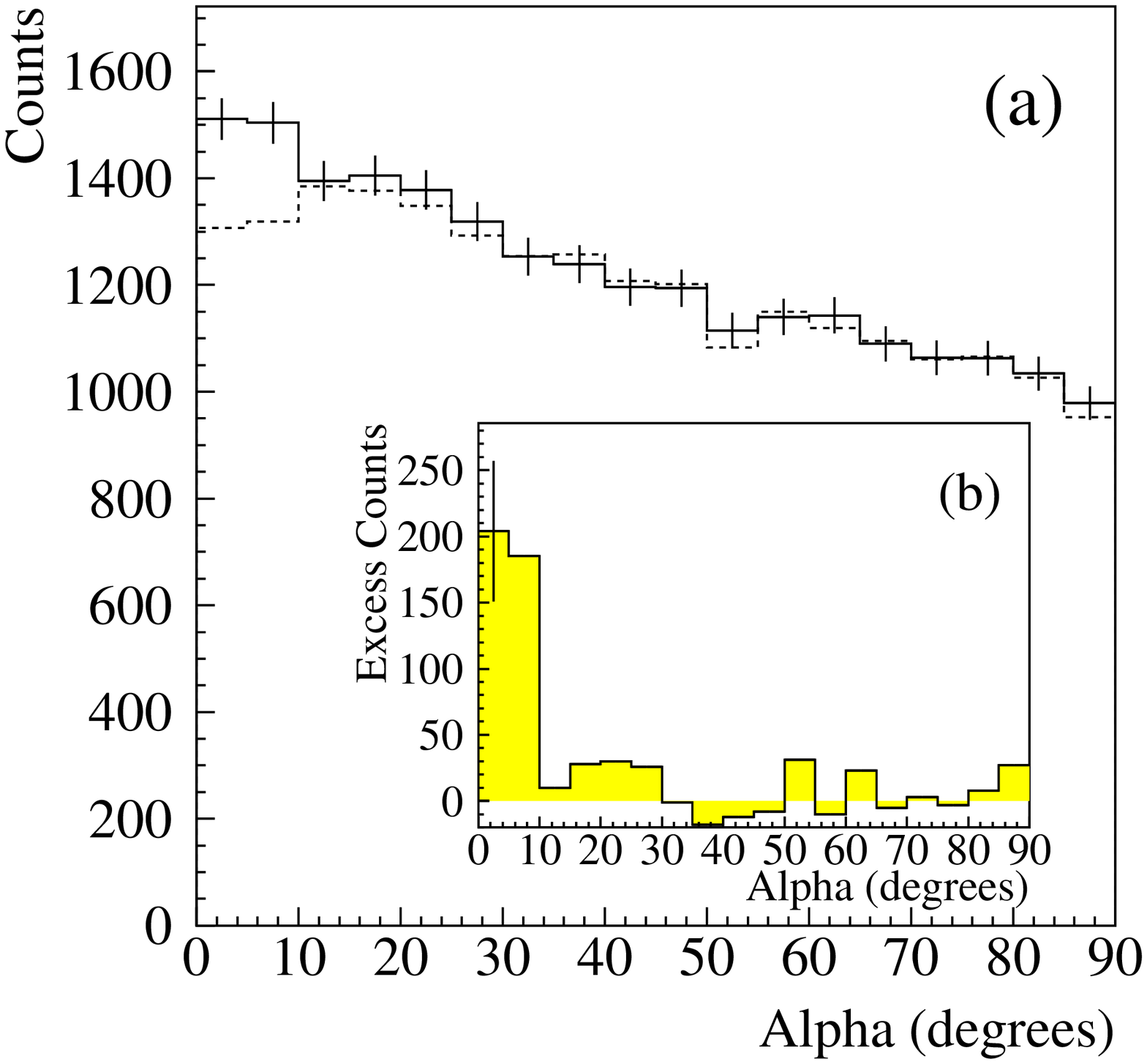]{Distributions of the image orientation angle
``alpha'' with respect to the direction of the maximum excess counts,
which is offset by about $0\fdg13$ to the southeast from the Vela pulsar
and corresponds to the peak position in Figure~3:
(a) on-source (solid line) and off-source (dashed line) distributions;
(b) distribution of the excess counts of the on-source above the background
(off-source) level.
For gamma-ray-like events, one expects this angle to be close to zero.
\label{fig1}}

\figcaption[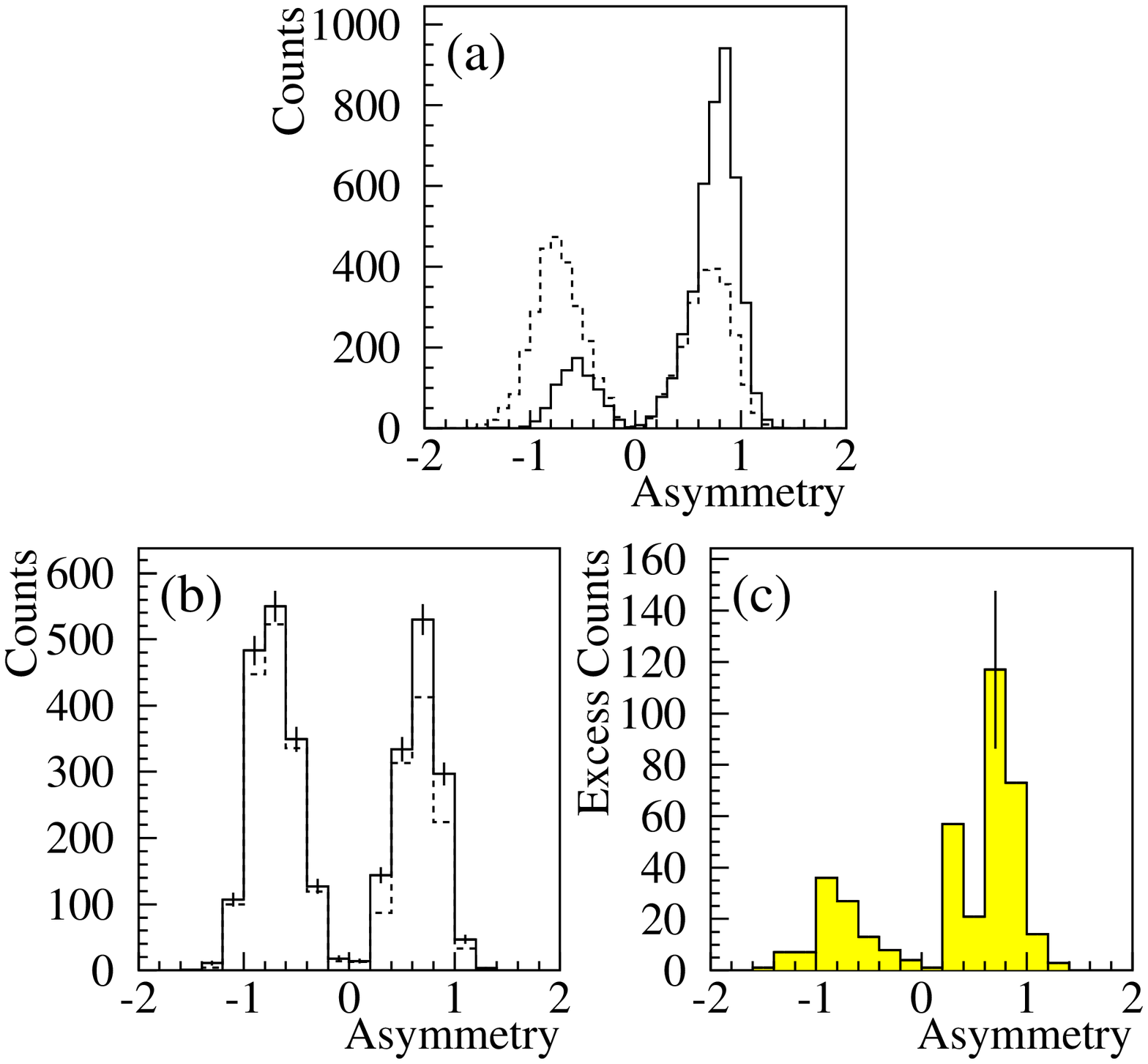]{Distributions of the ``asymmetry'' parameter
for simulated and observed \v{C}erenkov images:
(a) distributions of gamma-ray images for a point source (solid line)
and background (proton) images (dashed line) inferred from
Monte Carlo simulations;
(b) on-source (solid line) and off-source (dashed line) distributions
of gamma-ray-like events for the direction of the maximum excess counts,
which is the same as in Figure~1;
(c) distribution of the excess counts of the on-source above the background.
This parameter measures the pointing of the images toward the source,
as indicated by positive values. \label{fig2}}

\figcaption[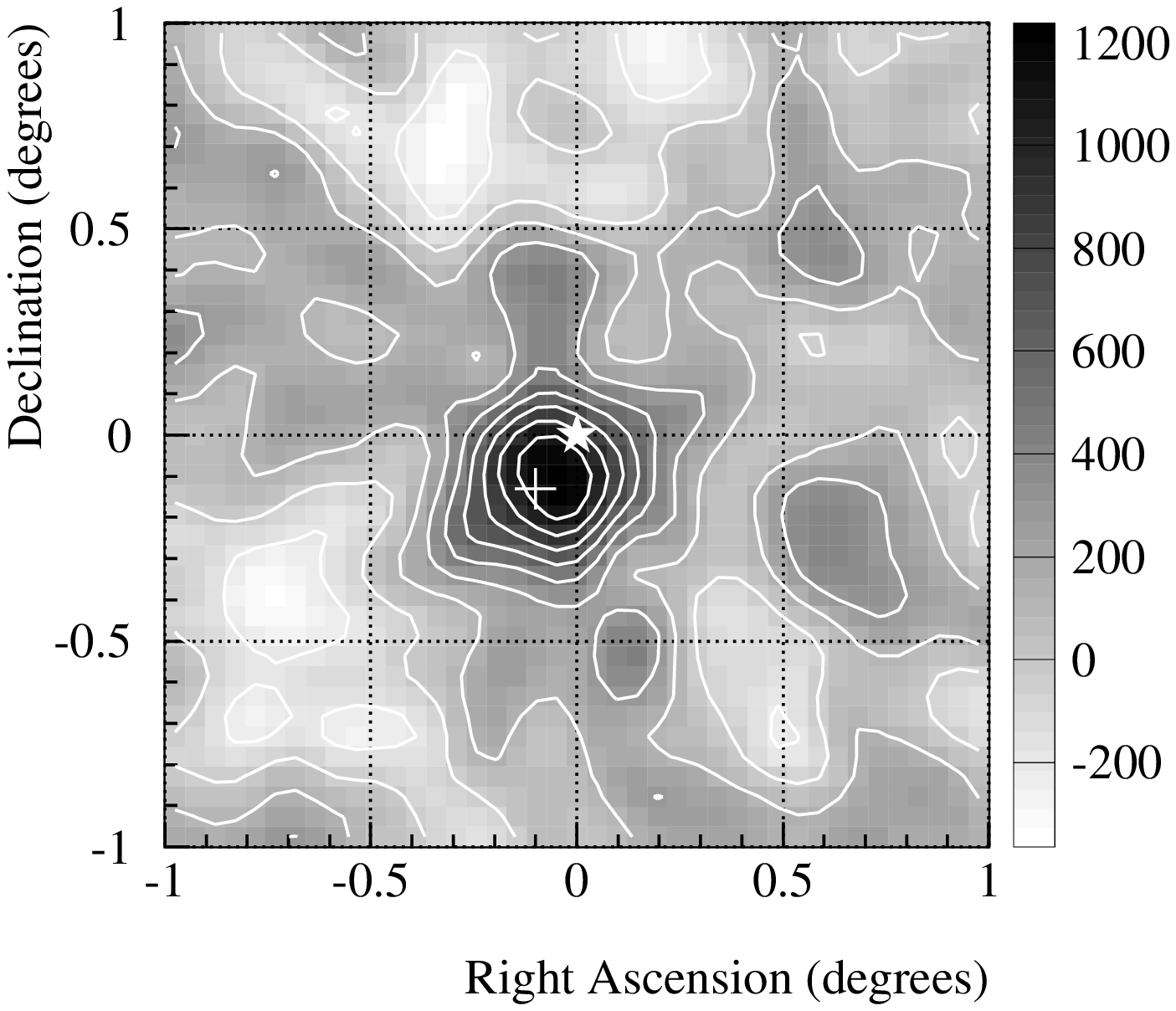]{A density map of excess counts around the Vela pulsar
plotted as a function of right ascension and declination; north is up and
east is to the left. The gray scale at the right is in
$\rm counts\; degree^{-2}$. The ``star'' at the origin of the map
indicates the position of the Vela pulsar. We have observed the same field
of view also in 1997 with a reduced energy threshold, to confirm the offset
of the source from the pulsar. The position of the maximum excess counts
from the 1997 data is indicated by the ``cross'' in the map. \label{fig3}}


\plotone{vela_alpha.eps}


\plotone{vela_asymmetry.eps}


\plotone{vela_map.eps}


\clearpage

\begin{references}
\reference{arzo1992} Arzoumanian, Z., Nice, D. \& Taylor, J.H. 1992,
                     GRO/radio timing data base, Princeton University
\reference{bail1989} Bailes, M., {\it et al.} 1989, \apjl, 343, L53
\reference{bhat1987} Bhat, P.N., {\it et al.} 1987, \aap, 178, 242
\reference{bowd1993} Bowden, C.C.G., {\it et al.} 1993, Proc. 23rd Int.
                     Cosmic Ray Conf. (Calgary), 1, 294
\reference{bucc1983} Buccheri, R., {\it et al.} 1983, \aap, 128, 245
\reference{deja1996} de~Jager, O.C., Harding, A.K. \& Strickman, M.S. 1996,
                     \apj, 460, 729
\reference{edwa1994} Edwards, P.G., {\it et al.} 1994, \aap, 291, 468
\reference{grin1975} Grindlay, J.E., {\it et al.} 1975, \apj, 201, 82
\reference{hara1993} Hara, T., {\it et al.} 1993, Nucl. Inst. Meth. Phys.
                     Res. A, 332, 300
\reference{hard1997} Harding, A.K. \& de~Jager, O.C. 1997, Proc.
                     32nd Rencontres de Moriond, in press
\reference{harn1985} Harnden, F.R., {\it et al.} 1985, \apj, 299, 828
\reference{hill1985} Hillas, A.M. 1985, Proc. 19th Int. Cosmic Ray Conf.
                     (La Jolla), 3, 445
\reference{kanb1994} Kanbach, G., {\it et al.} 1994, \aap, 289, 855
\reference{kawa1996} Kawai, N. \& Tamura, K. 1996, Proc. IAU Colloquium 160,
                     ``Pulsars: Problems and Progress'', ed. S.~Johnston,
                     M.A.~Walker and M.~Bailes, p.~367
\reference{kenn1984} Kennel, C.F. \& Coroniti, F.V. 1984, \apj, 283, 694
\reference{kifu1995} Kifune, T., {\it et al.} 1995, \apjl, 438, L91
\reference{kifu1996} Kifune, T. 1996, Proc. IAU Colloquium 160,
                     ``Pulsars: Problems and Progress'', ed. S.~Johnston,
                     M.A.~Walker and M.~Bailes, p.~339
\reference{mark1995} Markwardt, C.B. \& \"{O}gelman, H. 1995, \nat, 375, 40
\reference{mark1997} Markwardt, C.B. \& \"{O}gelman, H. 1997, \apjl, 480, L13
\reference{nel1993}  Nel, H.I., {\it et al.} 1993, \apj, 418, 836
\reference{ogel1989} \"{O}gelman, H., Koch-Miramond, L. \& Auri\'{e}re, M.
                     1989, \apjl, 342, L83
\reference{ogel1993} \"{O}gelman, H., Finley, J.P. \& Zimmermann, H.U. 1993,
                     \nat, 361, 136
\reference{prot1987} Protheroe, R.J. 1987, Proc. Astron. Soc. Aust., 7 (2), 167
\reference{punc1993} Punch, M. 1993, Ph.D. thesis, National University
                     of Ireland
\reference{rama1995} Ramanamurthy, P.V., {\it et al.} 1995, \apjl, 447, L109
\reference{reyn1993} Reynolds, P.T., {\it et al.} 1993, \apj, 404, 206
\reference{stan1982} Standish, E.M. 1982, \aap, 114, 297
\reference{tani1994} Tanimori, T., {\it et al.} 1994, \apjl, 429, L61
\reference{tayl1989} Taylor, J.H. \& Weisberg, J.M. 1989, \apj, 345, 434
\reference{vaca1991} Vacanti, G., {\it et al.} 1991, \apj, 377, 467
\reference{week1989} Weekes, T.C., {\it et al.} 1989, \apj, 342, 379
\reference{yosh1996} Yoshikoshi, T. 1996, Ph.D. thesis, Tokyo Institute of
                     Technology
\end{references}
\end{document}